\documentclass[aps,prl,preprint,groupedaddress]{revtex4-1}
\usepackage{graphicx}

\begin{document}


\title{A Quinone Based Single-Molecule Switch as Building Block for 
  Molecular Electronics}


\author{Herbert Fr{\"u}chtl}
\email[]{herbert.fruchtl@st-andrews.ac.uk}
\author{Tanja van Mourik}
\affiliation{EaStCHEM, School of Chemistry, University of St Andrews}


\date{\today}

\begin{abstract}
Azophenine has previously been identified as a controllable molecular 
switch when 
deposited on a Cu(110) surface, where it can be in two symmetry-equivalent 
states. Each of the two states can be set as well as read by means
of a scanning tunneling microscope (STM). We propose a family of molecules
based on the same quinone core, which show similar switching behavior without
a supporting metal 
surface. Such a molecule could be an element in a molecular circuit or computer.
Using the example of a simple hypothetical molecule, we show that it is possible
to create molecules that show the necessary properties: two conformations
with similar energy but different electric conductivity, and the possibility to
switch between those by applying an external electric field.
\end{abstract}


\maketitle

\section{Introduction}
While the miniaturization of semiconductor-based circuits is still following 
Moore's Law, which describes a doubling of transistors per surface area
every 18 months, this trend must reach its physical limits in the foreseeable 
future. One possible avenue of overcoming this barrier is the use of 
molecular electronics, where individual molecules would act as the building
blocks of electronic devices, such as transistors or memory elements. A
recent review article by Zhang {\it et al.} \cite{Zhang2015} demonstrates an
active research area.

Schaub {\it et al.} \cite{Renald_switch,RenaldSwitch2} have recently reported a
controllable switch
consisting of an azophenine molecule deposited on a Cu-(110) surface. If a
voltage greater than 0.3 V is applied, one of two symmetry-related
tautomers can be produced, depending
on the position of an STM tip. A smaller voltage allows the current
tautomeric state of the molecule to be determined without changing it.
Translated
into the language of computing, this constitutes a memory element that
can be written and read. 
Unfortuntely, the need for an STM tip to be moved to the correct position
above the molecule precludes the operation at a frequency that might be 
competitive with current microelectronics. An additional problem is that
changes in conductance are only
relevant in the direction perpendicular to the surface, because the supporting 
metal would short-circuit any voltage parallel to the surface.

In order to create a molecule that can be used in electronic devices, three
prerequisites have to be in place: bistability, {\it i.e.} the existence of two
states of the same or similar energy, a means of forcing the molecule to 
switch between these states, and a significant difference in conductance 
between those conformers. We will demonstrate that for a simple model system
based on the same quinone core, with a coordinated iron atom,
a similar amino-imino tautomerization mechanism as reported
in the azophenine switch can be observed. This makes it
possible to address all three of these requirements.

\begin{figure}
\includegraphics[width=\textwidth]{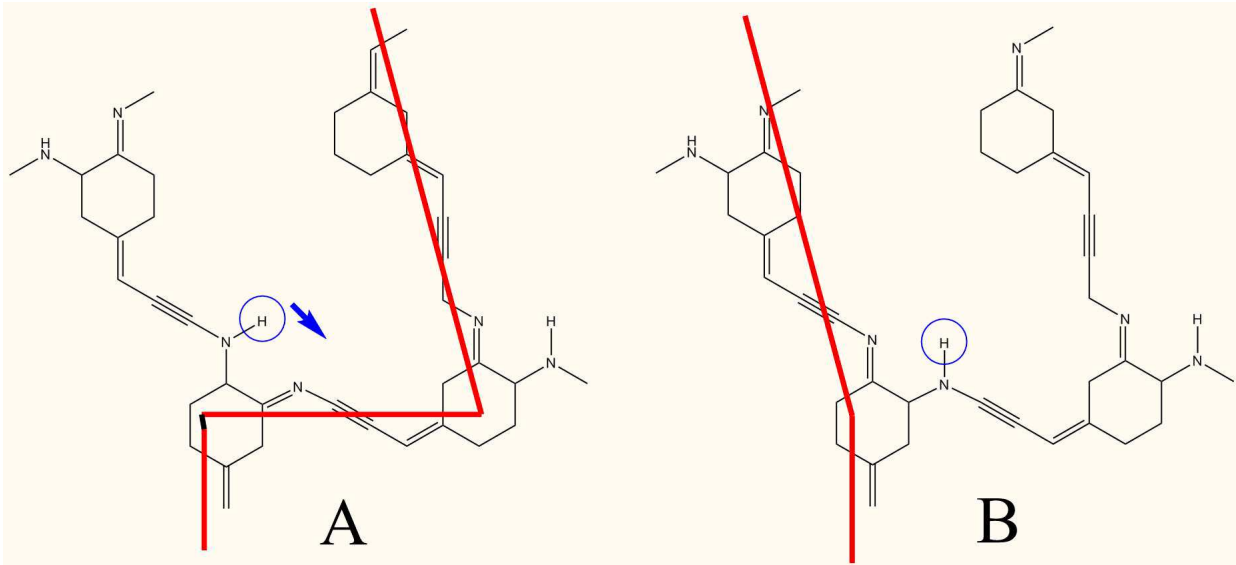}
\caption{\label{Fig:switch-network}
  Network of molecular switches, showing the change in flow current
  (indicated by the solid lines), upon switching the circled hydrogen
  as indicated by the arrow}.
\end{figure}

If such a switch could be produced, it would be possible to create networks of
switches, connected by molecular wires, in two or three dimensions. Figure
\ref{Fig:switch-network} shows a possible example, using acetylene as a
molecular wire, and for the sake of simplicity a quinone molecule with only
one imino-amino switch. Assuming that the imino double bond is conducting and
the amino single bond insulating, after switching the circled hydrogen atom as
indicated by the arrow, the current will flow from the left bottom ring to
the left upper ring, instead of from the left lower ring to the right
upper ring.

\section{Methodology}
The molecular geometry and enery profile were determined using the PBE 
\cite{PBE} hybrid functional with Grimme D3 dispersion correction and Becke-Johnson
damping \cite{BeckeD3,BJdamping}. The def2-TZVP basis set \cite{def2TZVP} was used for
all calculations. As the
aim of this study is a proof-of-principle on a model system, this level of
theory was considered sufficient. All calculations were carried out using
the ORCA \cite{ORCA} program.

The transition path was determined using a relaxed potential energy scan,
reducing the distance of the switching hydrogen to the (initially imino)
nitrogen from 2.2 {\AA} to 0.9 {\AA} in steps of 0.05 \AA. As the energy changes
more quickly in the later part of this scan, we reduced the step size to
0.025 {\AA} in the range below 1.5 \AA.
The energies with an applied electric field were then
calculated at the geometries determined during this scan. The
molecule with a coordinated Fe atom was treated as an open-shell triplet, as
this resulted in the lowest energy. Without a metal present, the molecule was
treated as closed shell. Where an electric field was applied, this was
oriented along the bond between the two carbon atoms furthest away from
the switching hydrogen. This will be generally parallel to the switching
direction, but the geometry of the carbon atoms defining the direction
of the current will be less prone to geometric changes during the
transition than atoms closer by. We did not take into account any changes in
geometry that might be caused by the electric field.

\section{Results and Discussion}
\begin{figure}[!htb]
\includegraphics[width=\textwidth]{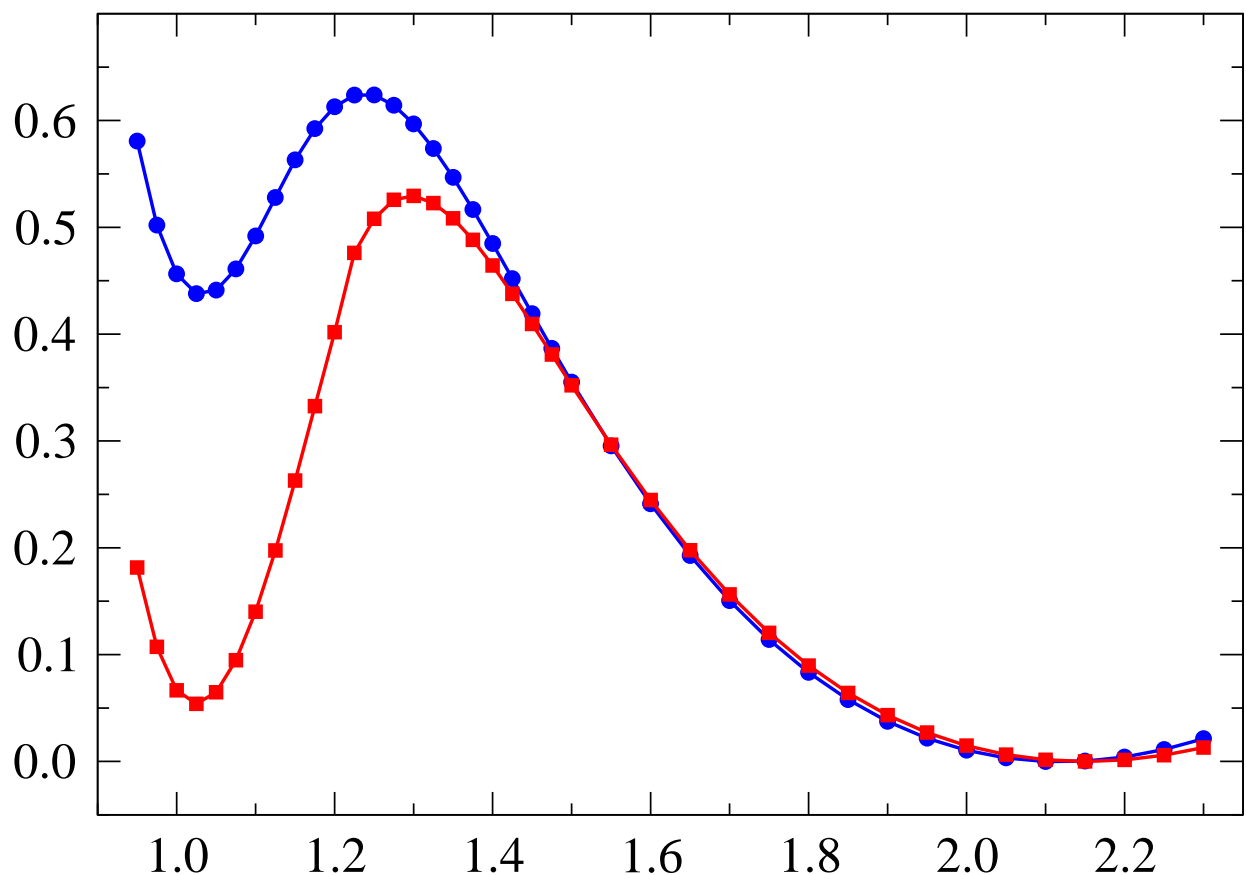}
\caption{\label{Fig:barriers-neutral} Energy profile along the switching
coordinate for the molecule with and without a coordinated Fe.}
\end{figure}

As discussed in the Introduction, we need to demonstrate the existence of
two conformations of similar energy. This can be achieved by coordinating an
iron atom above the quinone ring. Figure \ref{Fig:barriers-neutral} shows the
tautomerization barrier with and without this coordinated atom. Without a
metal atom, the quinone-like structure is, as expected, much more stable than
the ``switched'' alternative. The latter is metastable, but with a barrier
of 0.86 eV to reach, and a much smaller barrier (0.19 eV) for the backward
transition.
Note that the asymmetry in steepness of the curves, with the transition state
appearing closer to the minimum at small N-H distance, is an artifact of the choice of reaction
coordinate.

With the coordinated Fe atom, the difference is reduced to 0.05 eV, which is
much smaller than the barrier height. We believe that this difference in energy
between the two minima may be further reduced by changes in the molecule or its
surroundings. Thus, the first requirement for a molecular switch is fulfilled.

The second criterion is a considerable difference in conductivity between the
two tautomers along the chosen bonds. This can be assumed to be the case, since
the imino bond has double bond character and can function as the starting point
of a molecular wire consisting of conjugated double or triple bonds, such as,
e.g., $[-CH-]_{n}$ or $[C_2]_n$, which are known to be conducting,
whereas the amino
nitrogen is connected to both of its neighbors via single bonds, which will
break the conduction path.

The final requirement for a useable molecular switch is a means to trigger it.
We tested the effect of an external electric field on the energy barrier.
This would open up the opportunity to ``clock'' an electronic device
through, for
example, an oscillating external capacitor. We repeated our calculation of
the energy barrier in the presence of a field of 1 V/{\AA} in the positive and
negative direction along the switching path as described in the Methodology
section.

\begin{figure}
\includegraphics[width=\textwidth]{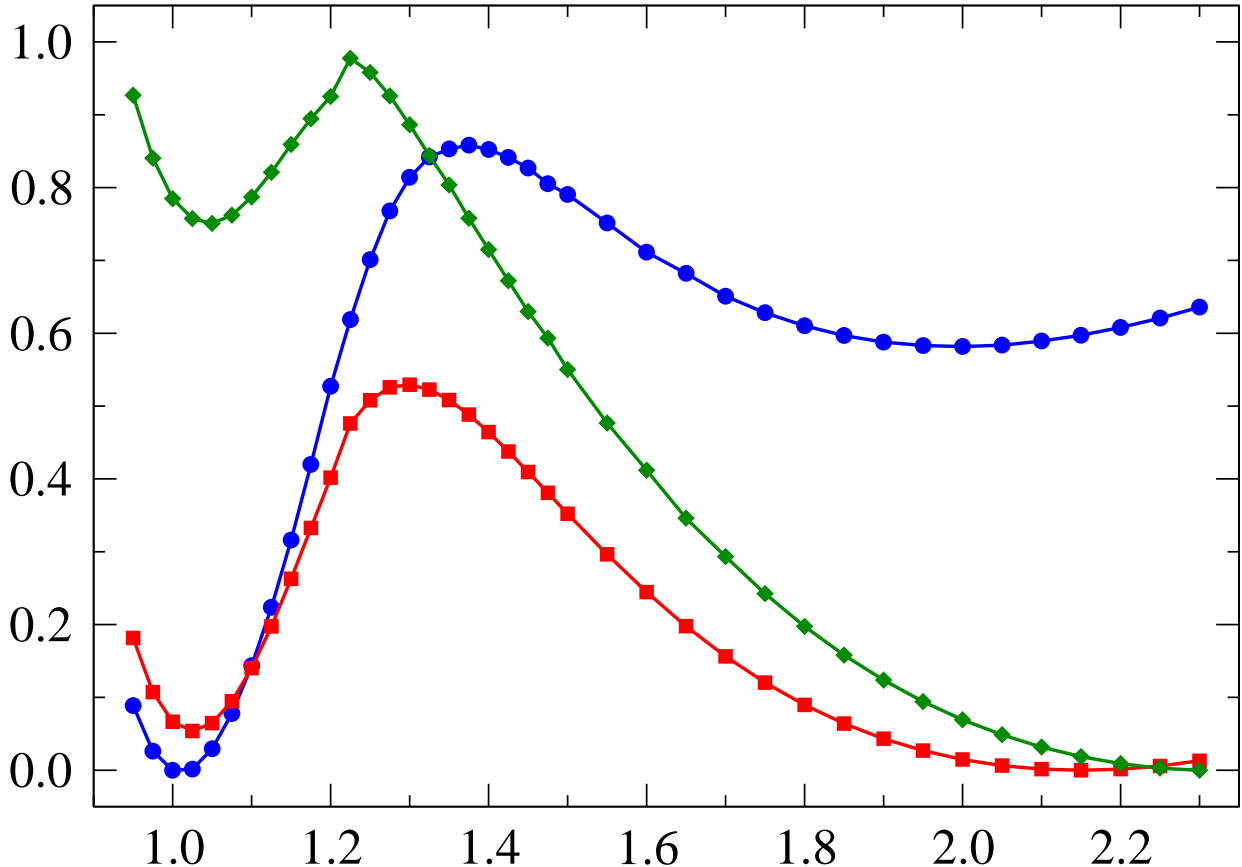}
\caption{\label{Fig:barriers-efield} Energy profile along the switching
coordinate in the presence of an external electric field.}
\end{figure}

Figure \ref{Fig:barriers-efield} shows the resulting barriers
compared to the neutral case. All curves are calibrated to show their lowest
point at zero.
While not removing the barriers completely, an electric field
introduces considerable asymmetry in the direction of the field. We have a
difference of more than 0.5 eV between the minima (0.75 eV and 0.58 eV
in the negative and
positive direction, respectively), a large
barrier of roughly 0.9 eV (0.86 eV and 0.98 eV for a negative and positive
field, respectively) in the unfavored direction
and less than a third of this (0.28 eV and 0.23 eV, respectively) towards the
lower-energy minimum.
While this asymmetry may not be sufficient to trigger an
immediate transfer of hydrogen in case of a reversal of the electric field
for our example molecule,
it indicates that such a scenario should be possible, either by using slightly
larger fields or lowering the barrier through a change in the chemical
environment, such as electron donating or withdrawing side groups.

\section{Conclusion and Outlook}
In conclusion, the current work shows that a non-surface-based
molecular switch satisfying our three criteria for good switching behavior
is in principle feasible.

It should be pointed out that this is a proof-of-principle study on a
model system. In a realistic
experiment, a means would have to be found to keep the coordinated metal atom
in place, such as caging it in surrounding organic groups, or sandwiching
it between two rings. A better understanding of the electronic structure
features enabling the switch with a coordinated metal atom may 
facilitate the design of other molecules with similar properties but without
the need for metal atoms.

If a network of switches was to be built, it would
initially need to be constructed in two dimensions on a non-conducting
surface or built into the third dimension as an organic or metal-organic
framework. Besides an electric field, as explored in the current Letter,
other switching mechanisms may include electronic excitations through
photons or the current through the molecule. In the latter case, the network
would be self-modifying and it would be possible to program various
types of logic into it.

\subsection{}
\subsubsection{}

\bibliography{herbert}

\end{document}